\documentclass[floatfix,twocolumn,prl,showpacs]{revtex4}
\usepackage{graphicx}
\usepackage{subfigure}
\usepackage{amssymb,amsmath}

\newcommand{\coloneq}{\; \colon \mspace{-12.0mu} =}

\begin{document}

\title{Dumb-bell swimmers}

\author{G. P. Alexander and J. M. Yeomans}
\affiliation{Rudolf Peierls Centre for Theoretical Physics, University of Oxford, \\
1 Keble Road, Oxford, OX1 3NP, England.}  

\date{\today}


\begin{abstract}
We investigate the way in which oscillating dumb-bells, a simple microscopic model of apolar swimmers, move at low Reynold's number. In accordance with Purcell's  Scallop Theorem a single dumb-bell cannot swim because its stroke is reciprocal in time. However the motion of two or more dumb-bells, with mutual phase differences, is not time reversal invariant, and hence swimming is possible. We use analytical and numerical solutions of the Stokes equations to calculate the hydrodynamic interaction between two dumb-bell swimmers and to discuss their relative motion. The cooperative effect of interactions between swimmers is explored by considering first regular, and then random arrays of dumb-bells. We find that a square array acts as a micropump. The long time behaviour of suspensions of dumb-bells is investigated and compared to that of model polar swimmers.
\end{abstract}

\maketitle

\section{Introduction}
\label{sec:intro}

Bacteria swim in a very different manner to macroscopic animals because, at micron length scales, inertial effects are negligible compared to viscous forces. In this, the zero Reynolds number limit, the Navier-Stokes equations reduce to the Stokes equations, which are time reversal invariant. This leads immediately to the Scallop Theorem~\cite{purcell77}, that any swimming stroke must be non-reciprocal -- look different when run forwards or backwards in time -- if it is to result in a net motion. Examples of strategies developed by microorganisms to overcome the Scallop Theorem are rotating flagella, waving cilia and surface waves. Moreover initial experiments have shown that it is possible to design tiny swimming robots which may be developed for drug delivery or manipulating payloads in microchannels~\cite{dreyfus05}.

To gain insight into the generic behaviour of microscopic swimmers several model systems have been defined and studied both analytically and numerically. These include squirmers~\cite{ishikawa06,ishikawa08}, Purcell's three link swimmer~\cite{purcell77,becker03,tam07}, three linked spheres~\cite{ramin,chris07} and the `pushmepullyou' swimmer~\cite{avron05}. All of these systems are polar swimmers, in that they undergo a non-reciprocal stroke, which picks out a preferred direction in time and hence results in swimming. Indeed, the Scallop Theorem suggests that apolar models should be immediately discounted as they will not swim.

However, as pointed out by Koiller {\it et al}~\cite{koiller96}, although this is true for a single apolar swimmer, a collection of such swimmers {\em can} swim. If their swimming strokes are not in phase, the motion taken as a whole is not reciprocal in time, and motility is possible. The physical mechanism which turns the motion of an apolar swimmer into a net displacement is the hydrodynamic forces between the swimmers.  

In this paper we study the simplest model of an apolar swimmer, oscillating dumb-bells. We give an analytic solution for the motion of a pair of dumb-bells, in the Oseen tensor limit, and valid for large separations. By numerically iterating these results we are able to gain insight into how the orbits of two swimmers depend on their relative positions, separations, and phases. We then consider regular arrays of swimmers and, in particular, suggest a way in which tethered swimmers could be used as a micropump. 

We next present results for suspensions where the dumb-bells are initially placed randomly with random relative phases. The collective behaviour of such groups of self-propelled organisms has received a lot of interest in the past decade. Remarkably, the dynamics of systems as seemingly unrelated as flocks of birds, swarms of bacteria and vibrated rods can be treated within the common theoretical framework of active fluids~\cite{toner05,ramaswamy06}. Within this class a distinction has been made between systems where the order is polar, for example, flocks of birds, and those with apolar order, such as vibrated rods.

Since a dumb-bell is intrinsically apolar, a suspension of them provides a minimal microscopic model of an active {\em apolar} fluid. We present results comparing the behaviour of a suspension of interacting dumb-bells to that of a suspension of  three-linked-sphere swimmers~\cite{ramin,chris07}, which is a simple microscopic model of an active {\em polar} fluid.

\section{Hydrodynamic interactions between oscillating dumb-bells}
\label{sec:twodipoles}

We first describe the motility of oscillating dumb-bells at zero Reynolds number. Each dumb-bell comprises two spheres, of radius $a$, joined by a thin, rigid rod, whose length varies sinusoidally as $D+\xi \sin (\omega t)$. We stress the importance of the relative phase of the two dumb-bells: if they oscillate in phase or $\pi$ out of phase then their combined movements remain reciprocal and the Scallop Theorem prevents any net motion~\cite{purcell77}. However, for other values of the relative phase, motion will occur. To show this we use the Oseen tensor formulation of hydrodynamics, valid in the limit of zero Reynolds number~\cite{happel65}. We present anaytic calculations, valid for large separations, and numerical results, valid for separations down to of order the size of a dumb-bell.

Linearity of the Stokes equations which govern zero Reynolds number flows allows the fluid velocity to be written as a linear combination of the forces, ${\bf f}$, acting on the fluid due to the motion of the swimmers
\begin{equation}
{\bf u}({\bf x}) = \sum_{A} \Bigl[ G^1_A({\bf x}){\bf f}^1_A + G^2_A({\bf x}){\bf f}^2_A \Bigr] \; .
\label{eq:flow}
\end{equation}
Here the subscript $A$ labels the dumb-bells and the superscripts $1,2$ label the spheres of an individual dumb-bell. For spheres whose radii $a$ is small compared to their separation the Greens function, $G^r_A({\bf x})$, may be taken to be the Oseen tensor, 
\begin{equation}
G^r_A({\bf x}) = \begin{cases} \tfrac{1}{6\pi \mu a} \; \text{Id} \; , \quad & \text{if} \; |{\bf y}| = a \; , \\ \tfrac{1}{8\pi \mu} \tfrac{1}{|{\bf y}|} \bigl[ \text{Id} + \hat{{\bf y}} \otimes \hat{{\bf y}} \bigr] \; , \quad & \text{otherwise,}  \end{cases}
\label{eq:oseentensor}
\end{equation}
where ${\bf y} \coloneq {\bf x} - {\bf x}^r_A$.
The motion of the dumb-bells through the fluid is then determined by three ingredients; consistency of the fluid flow with the change in shape, i.e., the change in rod length, the constraint that each dumb-bell is force free, ${\bf f}^1_A + {\bf f}^2_A = 0$, and the constraint that each dumb-bell is torque free, $({\bf x}^2_A - {\bf x}^1_A) \wedge {\bf f}^1_A = 0$. 

A short calculation then leads to an expression for the forces
\begin{equation}
\begin{split}
{\bf f}^1_A & = -3\pi \mu a \Bigl\{ \xi \omega \cos (\omega t) \Bigl( 1 + \tfrac{3}{2} \tfrac{a}{D + \xi \sin (\omega t)} \Bigr) {\bf n}_A \\ 
& \quad + \bigl( D + \xi \sin (\omega t) + \tfrac{3a}{4} \bigr) \Omega_A \wedge {\bf n}_A + \dots \Bigr\} \\
& \quad + 3\pi \mu a \sum_{B \not = A} \Bigl\{ G^1_B({\bf x}^2_A) - G^1_B({\bf x}^1_A) \\
& \qquad - G^2_B({\bf x}^2_A) + G^2_B({\bf x}^1_A) \Bigr\} {\bf f}^1_B \; ,
\end{split}
\label{eq:force}
\end{equation}
where terms of $O[(a/D)^2]$ have been omitted. The unit vector ${\bf n}_A$ gives the direction of sphere $2$ relative to sphere $1$ and thus describes the orientation of the dumb-bell. Applying the torque free constraint determines the angular velocity, $\Omega_A$, from which we obtain an equation for the evolution of the dumb-bell's orientation 
\begin{align}
\frac{\text{d}{\bf n}_A}{\text{d}t} & = \Omega_A \wedge {\bf n}_A \; , \\
\begin{split}
& = \tfrac{1}{D + \xi \sin (\omega t)} \Bigl[ \text{Id} - {\bf n}_A \otimes {\bf n}_A \Bigr] \\
& \quad \times \sum_{B \not = A} \Bigl\{ G^1_B({\bf x}^2_A) - G^1_B({\bf x}^1_A) \\
& \qquad - G^2_B({\bf x}^2_A) + G^2_B({\bf x}^1_A) \Bigr\} {\bf f}^1_B \; .
\end{split}
\label{eq:rotate}
\end{align}
Finally, the translational motion of each dumb-bell is given by 
\begin{equation}
\begin{split}
\frac{\text{d}{\bf x}^c_A}{\text{d}t} & = \frac{1}{2} \sum_{B \not = A} \Bigl\{ G^1_B({\bf x}^2_A) + G^1_B({\bf x}^1_A) \\
& \quad - G^2_B({\bf x}^2_A) - G^2_B({\bf x}^1_A) \Bigr\} {\bf f}^1_B \; ,
\end{split}
\label{eq:translate}
\end{equation}
where ${\bf x}^c_A$ denotes the `centre' of the dumb-bell, ${\bf x}^c_A \coloneq ({\bf x}^1_A + {\bf x}^2_A)/2$. 
As expected, the motion described by Eqns.~\eqref{eq:rotate} and~\eqref{eq:translate} arises solely through interactions with other dumb-bells.

We assume that the dumb-bells are in a dilute suspension so that the separation, $r$, of any given dumb-bell from its nearest neighbour may be assumed to be large compared to its size, $D$. Under such circumstances the interactions between the dumb-bells may be expanded in a power series in $(D/r)$ and only the leading contributions retained. 
Integrating over a complete cycle leads to expressions for the changes in position and orientation of the dumb-bells after a single swimming stroke:
\begin{equation}
\begin{split}
\Delta {\bf n}_A & = \sum_{B \not = A} \frac{15a {\cal A}}{32\; r^5_{BA}} \Bigl( \hat{{\bf r}}_{BA} - ({\bf n}_A \cdot \hat{{\bf r}}_{BA}){\bf n}_A \Bigr) \\ 
& \; \times \Bigl\{ 3({\bf n}_A \cdot \hat{{\bf r}}_{BA}) + 6({\bf n}_A \cdot {\bf n}_B)({\bf n}_B \cdot \hat{{\bf r}}_{BA}) \\ 
& \;\; + 6({\bf n}_A \cdot {\bf n}_B)^2({\bf n}_A \cdot \hat{{\bf r}}_{BA}) - 7({\bf n}_A \cdot \hat{{\bf r}}_{BA})^3 \\
& \;\; - 21({\bf n}_A \cdot \hat{{\bf r}}_{BA})({\bf n}_B \cdot \hat{{\bf r}}_{BA})^2 \\
& \;\; - 42({\bf n}_A \cdot {\bf n}_B)({\bf n}_A \cdot \hat{{\bf r}}_{BA})^2({\bf n}_B \cdot \hat{{\bf r}}_{BA}) \\
& \;\; + 63({\bf n}_A \cdot \hat{{\bf r}}_{BA})^3({\bf n}_B \cdot \hat{{\bf r}}_{BA})^2 \Bigr\} \; ,
\end{split} 
\label{eq:updaten}
\end{equation}
\begin{equation}
\begin{split}
\Delta {\bf x}^c_A & = \sum_{B \not = A} \frac{9a {\cal A}}{32\; r^4_{BA}} \Bigl\{ \hat{{\bf r}}_{BA} \Bigl[ 1 + 2({\bf n}_A \cdot {\bf n}_B)^2 \\
& \;\; - 5({\bf n}_A \cdot \hat{{\bf r}}_{BA})^2 - 5({\bf n}_B \cdot \hat{{\bf r}}_{BA})^2 \\
& \;\; - 20({\bf n}_A \cdot {\bf n}_B)({\bf n}_A \cdot \hat{{\bf r}}_{BA})({\bf n}_B \cdot \hat{{\bf r}}_{BA}) \\
& \;\; + 35({\bf n}_A \cdot \hat{{\bf r}}_{BA})^2({\bf n}_B \cdot \hat{{\bf r}}_{BA})^2 \Bigr] \\
& \; + 2{\bf n}_A \Bigl[ ({\bf n}_A \cdot \hat{{\bf r}}_{BA}) + 2({\bf n}_A \cdot {\bf n}_B)({\bf n}_B \cdot \hat{{\bf r}}_{BA}) \\
& \;\; - 5({\bf n}_A \cdot \hat{{\bf r}}_{BA})({\bf n}_B \cdot \hat{{\bf r}}_{BA})^2 \Bigr] \Bigr\} \; ,
\end{split}
\label{eq:updatex}
\end{equation} 
where ${\bf r}_{BA}$ is the position vector of $B$ relative to $A$. The amplitude ${\cal A}$ is given by
\begin{equation}
{\cal A} = \pi D^2 \xi_A \xi_B \sin (\eta_{BA}) + \tfrac{\pi}{8} \xi^2_A \xi^2_B \sin (2\eta_{BA}) \; ,
\end{equation}
where $\eta_{BA}$ is the phase of $B$'s swimming stroke relative to $A$'s. The long time behaviour of a group of dumb-bells may be determined by numerically iterating Eqns.~\eqref{eq:updaten} and~\eqref{eq:updatex} to find the new positions and orientations of all the dumb-bells after each swimming cycle.

\section{Two dumb-bells}
\label{sec:twodbs}

\begin{figure}
\centering
\includegraphics[width=.4\textwidth]{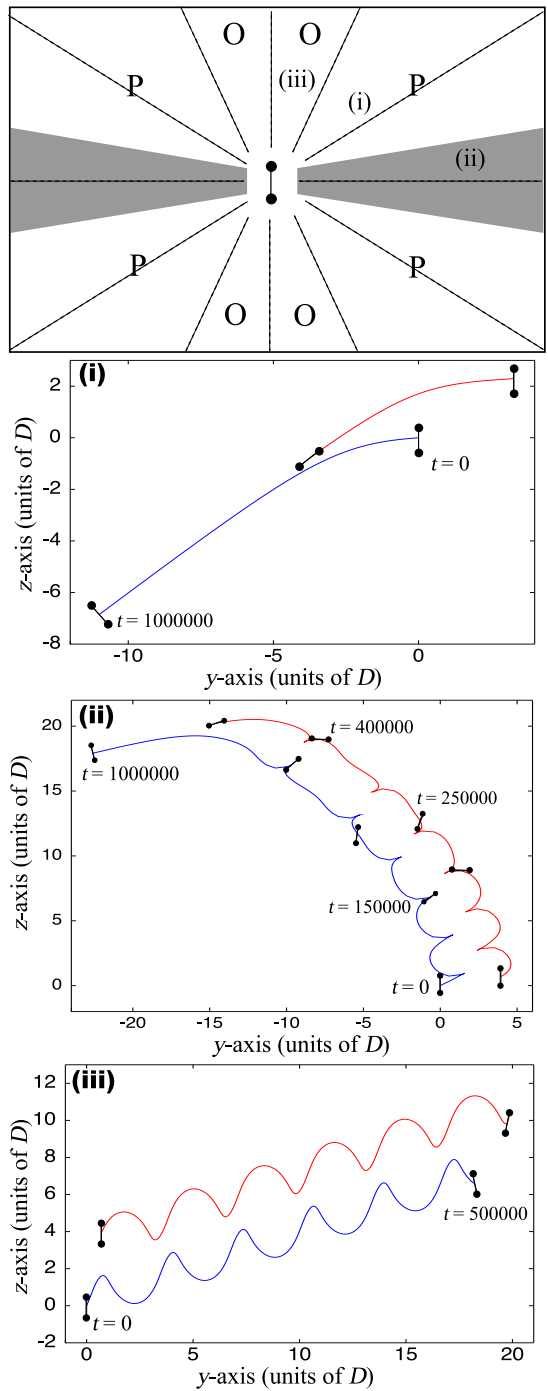}
\caption{(Colour online) Orbits of a pair of coplanar dumb-bells. The two dumb-bells are initially parallel with $A$ at the origin and $B$ placed at the corresponding point in the diagram. Several regimes of long time behaviour are found. In region P the two dumb-bells adopt a stable perpendicular configuration as exemplified by (i). This type of behaviour is also found in the shaded regions, however, before acquiring the perpendicular configuration the two dumb-bells tumble for a period of time, as shown in (ii). In the regions labelled O the dumb-bells move along parallel trajectories with an oscillatory motion, as illustrated in (iii). Finally, along each of the dashed lines there is no rotational interaction and the dumb-bells undergo a pure translation along straight parallel paths.}
\label{fig:pair}
\end{figure}

We first consider two dumb-bells lying in the $yz-$plane and both oriented along the $z-$direction. Dumb-bell $A$ is initially at the origin and dumb-bell $B$ is placed on a circle of radius $4D$ centred on the origin. For two dumb-bells, varying the relative phase does not lead to any qualitative changes in behaviour and therefore we consider only the case $\eta_{BA} = \pi/2$. The sole free parameter is the angle, $\theta$, that the position vector of $B$ makes with the $y-$direction. As this angle is varied the hydrodynamic interactions between the dumb-bells change leading to different long time behaviour, which we illustrate in Fig.~\ref{fig:pair}.

The predominant behaviour is for the two dumb-bells to adopt a perpendicular configuration in which one dumb-bell is oriented parallel and the other perpendicular to their relative position vector. This stable arrangement appears at long times for all initial configurations in the regions labelled P in Fig.~\ref{fig:pair}. An exemplary time series showing how the perpendicular configuration is reached is shown in Fig.~\ref{fig:pair}(i). The stability of this state may be seen from a linear stability analysis of Eq.~\eqref{eq:updaten}. 

Similarly, linear stability analysis reveals that the rotational fixed point at $\theta = 0$ is unstable. However, small deviations away from this fixed point do not lead smoothly to the stable perpendicular configuration. Instead there is an initial period during which the dumb-bells tumble, often several times, before they finally settle down, as illustrated in Fig.~\ref{fig:pair}(ii). Exactly at the fixed point, and in the absence of any fluctuations, the dumb-bells undergo a pure translational motion, swimming cooperatively in the direction of the dumb-bell with positive relative phase.

The fully aligned configuration with $\theta = \pi /2$ is also a rotational fixed point and again, exactly at this angle, the motion is purely translational with both dumb-bells moving in the same direction. However, in this case, the fixed point is a centre and small deviations away from it lead to an oscillatory cooperative motion, an exemplary time series of which is shown in Fig.~\ref{fig:pair}(iii). This oscillatory motion occurs throughout the regions labelled O in Fig.~\ref{fig:pair} and is separated from the P regions by an additional rotational fixed point at $\theta \approx 65^o$.

\section{Linear chains of dumb-bells}
\label{sec:chains}

We now present examples of the cooperative effects of the hydrodynamic interaction between many apolar swimmers, for first regular, and then random, distributions of swimmers. Consider a chain of $N$ identical dumb-bells all oriented along the $z-$direction and initially positioned at equally spaced intervals of $5D$ along the $y-$axis. To optimise the hydrodynamic interactions between nearest neighbours the relative phase between any two neighbouring dumb-bells is set to $\pi/2$, increasing in the positive $y-$direction. Although this is a highly artificial configuration, the lack of any rotation greatly simplifies the dynamics, allowing for the effect of changing $N$ to be clearly quantified.

\begin{figure}
\centering
\includegraphics[width=.47\textwidth]{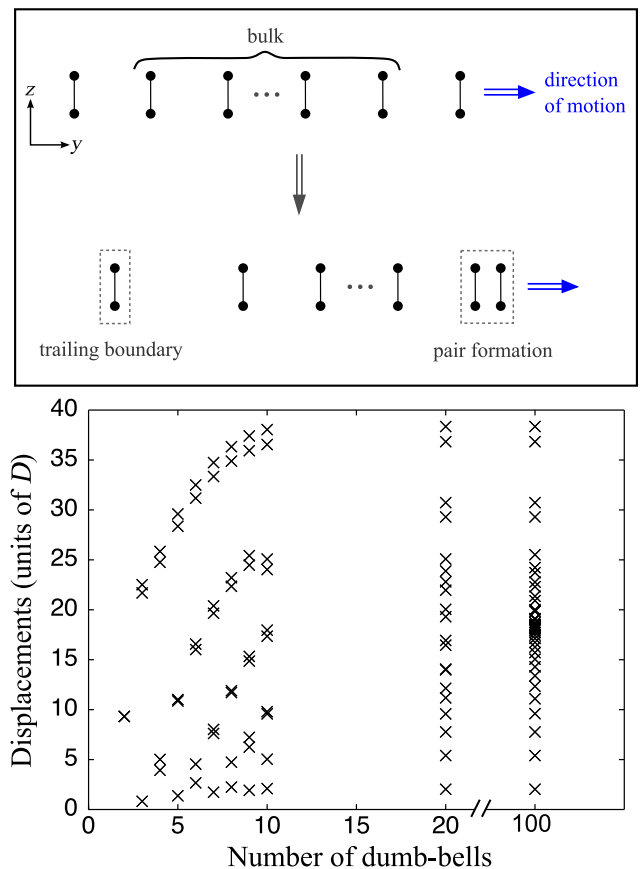}
\caption{(a) Configuration of a linear chain of dumb-bells and a schematic representation of its evolution. The bulk of the chain moves faster than the boundary, leading to dumb-bells being left behind at the trailing boundary and the formation of `fast pairs' at the leading boundary. (b) The distances moved by each dumb-bell in the chain after $100,000$ swimming strokes for chains with different numbers of dumb-bells.}
\label{fig:chainconfig}
\end{figure}

We show in Fig.~\ref{fig:chainconfig} the initial configuration of the chain and the distances moved by each of the dumb-bells as their total number increases. Two features are particularly noteworthy: firstly the behaviour for $N>2$ is significantly different from that for $N=2$, and secondly the general behaviour for $N>20$ shows only minor variations, indicating that an asymptotic limit is being approached.

The evolution of the pattern of swimmers may be understood by noting that there is a fundamental distinction between dumb-bells with two nearest neighbours and those with only one. The former constitute what we shall call the {\em bulk} of the chain, while the latter form the {\em boundary}. A dumb-bell in the bulk not only gets pulled along by its neighbour in front of it, but is also pushed by its neighbour from behind. These two interactions add constructively leading to a member of the bulk moving faster than a single isolated pair. By contrast a dumb-bell which is on the boundary only has one nearest neighbour and hence does not benefit from this added boost. Thus the bulk moves faster than the boundary.

This has the effect of introducing an asymmetry between the two boundaries; the dumb-bell on the trailing boundary gets left behind while that on the leading boundary is caught up. From Eq.~\eqref{eq:updatex} we see that the strength of interactions depends on the separation of the dumb-bells as $r^{-4}$, so that as the trailing dumb-bell gets left behind, the pull it receives from the bulk rapidly diminishes until it becomes isolated and can no longer move. At the same time the distance between the leading dumb-bell and the bulk decreases yielding a sharp increase in the strength of interaction between itself and its neighbour. As a result the leading pair speed up significantly and are ejected from the front of the chain.

\section{Lattice pumps}
\label{sec:groups}

The cooperative motion of one dimensional chains carries over to regular arrays in two dimensions. As an example, consider a square lattice of dumb-bells all oriented in the $z-$direction and with $(x,y)$ positions $(jL,kL)$, where $L$ is the lattice constant and $j,k$ are integers. Co-operative directed motion can be induced by defining the phase of each dumb-bell to be $\phi_{j,k} = \pi (j+k)/2$. The entire lattice then moves uniformly along the $[110]$ direction.

For a system of real swimmers this state could not be sustained; we have found that it is unstable to any imperfections of the lattice such as boundaries or fluctuations in the position of the swimmers.
This instability of long range coherent states is in agreement with analytic work by Ramaswamy and co-workers~\cite{simha02,toner05,ramaswamy06} and numerical simulations of Saintillan and Shelly~\cite{saintillan07}. However one might envisage tethering fabricated dipolar swimmers to a substrate and aligning and activating them with a magnetic field. Such a set-up would act as a micon-scale pump. We estimate $v \sim \mu m\, s^{-1}$, which is similar to the velocities achieved by bacteria. It should be noted that the interactions scale with separation as $r^{-4}$, so that a small decrease in lattice spacing will provide a substantial increase in flow speed.

\section{Suspensions of dumb-bells}
\label{sec:suspension}

\begin{figure}[t]
\centering
\includegraphics[width=.49\textwidth]{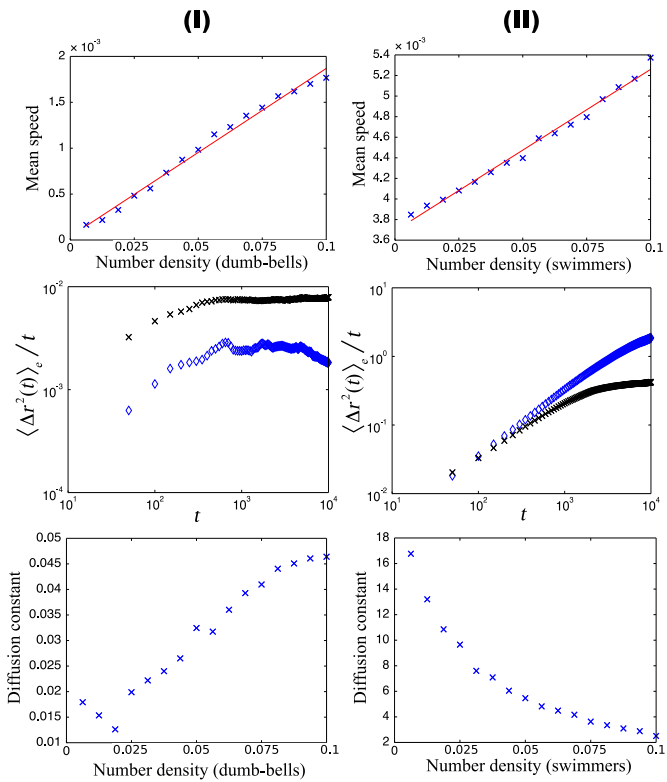}
\caption{(Colour online) Collective properties of a suspension of interacting (I) dumb-bell swimmers (apolar) and (II) three-sphere swimmers (polar). Top: mean speed of the swimmers as a function of their number density. The line indicates a linear fit. Middle: mean square displacement as a function of time for number densities 0.1 ($\times$) and 0.02 ($\diamond$). Bottom: diffusion constant for increasing number of swimmers.}
\label{fig:collective}
\end{figure}  

We consider a suspension of $N$ dumb-bells initially dispersed throughout a cubic box of side $L = 20D$ with random positions, orientations and relative phases. To avoid singularities in the hydrodynamic interactions a short distance cut-off is employed when the separation between any two dumb-bells becomes less than $0.5D$. At separations of order $D$ an expansion of the Oseen tensor in powers of $D/r$ does not converge rapidly. To overcome these difficulties we instead use an expansion in powers of $\xi/r$ to better describe the near field hydrodynamics.  

The combination of a random initial configuration and apolar symmetry means that there is no prefered direction for the motion, and on average the velocity is zero. In continuum models of apolar active fluids spontaneous symmetry breaking can lead to a state with non-zero average velocity~\cite{voituriez05}, however we have not observed any such transitions in our simulations of dumb-bells. In the absence of a net velocity, the mean speed provides a measure of the degree of collective activity. Fig.~\ref{fig:collective}(I) shows that the mean speed of the dumb-bells increases linearly with the number density $n=N/L^3$. This agrees with a simple scaling argument: since every dumb-bell will interact with every other one, the total number of interactions scales as $n^2$. Balancing this against the kinetic energy predicts that the mean speed should increase linearly with $n$. The same linear scaling, but tending to a finite value as $N \rightarrow 1$ because of the finite speed of a single swimmer,  is also found for a suspension of polar swimmers (Fig.~\ref{fig:collective}(II)).

A measure of the nature of the collective motion generated by the interacting dumb-bells is the mean square displacement, $\langle \Delta r^2(t) \rangle_{e} = \langle \lvert {\bf r}(t) - {\bf r}(0) \rvert^2 \rangle_{e}$, where $\langle \cdot \rangle_{e}$ denotes an ensemble average. This is plotted in Fig.~\ref{fig:collective} for number densities of $n=0.02$ and $n=0.1$. At long times, and for large number densities, a scaling form $\langle \Delta r^2(t) \rangle_{e} \sim t^z$ develops, with an exponent of $z=1$ indicating that the suspension of dumb-bells is behaving diffusively. At smaller number densities the behaviour is more sporadic because the large average separation between dumb-bells greatly reduces the strength of the interactions between them. The collective motion is then dominated by those fluctuations in the local number density which bring two dumb-bells close enough to allow them to move appreciably.

The mean square displacement of a suspension of dumb-bells shows qualitatively different behaviour to that of a suspension of polar swimmers. In the latter case, shown in Fig.~\ref{fig:collective}(II), the mean square displacement is ballistic at short times ($z=2$), with a cross-over to diffusion ($z=1$) at longer times. Moreover, since this cross-over is due to the randomisation of swimmer orientations through hydrodynamic interactions, it occurs at later and later times as the number density is reduced. For polar swimmers this leads to a diffusion constant which decreases as the number density is increased. For apolar swimmers the oppposite is true; their motion arises solely from hydrodynamic interactions and the diffusion constant increases with increasing number density.

\section{Discussion}
\label{sec:discuss}

The aim of this letter is to discuss the motion of a simple model of apolar swimmers, systems which move at zero Reynolds number {\em only} in the presence of other swimmers. We have demonstrated that, as is the case for colloids~\cite{janosi97} and polar swimmmers~\cite{chris07}, hydrodynamic interactions lead to complex collective behaviour. 

Particular observations are that, for two dumb-bell swimmers which are initially parallel, the most likely final state corresponds to the axes of the swimmers lying at right angles. Oscillatory trajectories are also observed. For regular arrays of swimmers cooperation between hydrodynamic pair interactions can lead to simple flow fields. In particular a square array of dumb-bells produces a constant flow and hence, if the dumb-bells were fixed in position, could act as a pump. For a large number of dumb-bell swimmers, initialised with random positions and phases, the mean speed is linear in the number of dumb-bells, reflecting the energy pumped into the system by the hydrodynamic interactions. The  mean-square displacement evolves linearly in time showing the expected diffusive behaviour.

There is much further work to the done to explore the phase space of apolar swimmers, for example considering initial configurations for which pairs move out of the plane, three dumb-bell orbits, and the role of the distribution of phases on multi-dumb-bell motion. For more than two dumb-bells the relative phase becomes an important variable; since the pairwise interactions between three or more dumb-bells cannot simultaneously take their maximum value the system exhibits a type of frustration. It is also interesting to consider the effect of moving away from the zero Reynolds number limit~\cite{lauga07}, and to ask which of the properties of dumb-bells provide a generic representation of the class of apolar swimmers. Comparing simple microscopic models of polar and apolar swimming may help to formulate the correct continuum theory of swimmers, and to link the microscopic and continuum length scales.

\acknowledgements

We are grateful to Mike Cates, Davide Marenduzzo and Chris Pooley for useful discussions and thank Eric Lauga and Denis Bartolo for showing us reference~\cite{lauga08} prior to publication.

\end{document}